\documentclass[fleqn,usenatbib]{mnras}

\usepackage{newtxtext,newtxmath}
\usepackage[T1]{fontenc}
\usepackage[font=small,labelfont=bf,tableposition=top]{caption}

\usepackage[T1]{fontenc}
\usepackage{ae,aecompl}

\usepackage{myaasmacros}

\usepackage{booktabs,tabularx,graphicx,amsmath,wasysym,makecell,hhline,enumitem,fancyvrb,gensymb,multirow,ulem,pifont,adjustbox}
\defcitealias{Renaud2021}{Paper I}

\title[Stellar kinematics
in gas rich disc galaxies]{From giant clumps to clouds - II. The emergence of thick disc kinematics from the conditions of star formation in high redshift gas rich galaxies}

\author{Floor van Donkelaar et al. \\
$^1${\small Lund Observatory, Department of Astronomy and Theoretical Physics, Lund University, Box 43, SE-221 00 Lund, Sweden}\\
}

\author[Floor van Donkelaar et al.] 
{Floor van Donkelaar$^{1,2}$\thanks{floor.vandonkelaar@uzh.ch}, Oscar Agertz$^{1}$ and Florent Renaud$^{1}$\\
$^1$ Lund Observatory, Department of Astronomy and Theoretical Physics, Box 43, SE-221 00 Lund, Sweden \\
$^2$Center for Theoretical Astrophysics and Cosmology, Institute for Computational Science, University of Zurich, Winterthurerstrasse 190, CH-8057 Z\"urich, Switzerland }
\date{Accepted XXX. Received YYY; in original form ZZZ}

\pubyear{2021}

\begin{document}
\label{firstpage}
\pagerange{\pageref{firstpage}--\pageref{lastpage}}
\maketitle

\begin{abstract}
High redshift disc galaxies are more gas rich, clumpier, and more turbulent than local Universe galaxies. This early era of galaxy formation imprints the distribution and kinematics of the stars that we observe today, but it is not yet well established how. In this work, we use simulations of isolated Milky Way-mass disc galaxies to study how kinematic properties of stars change when varying the gas fraction. This allows us to quantify the roles played by internal processes, e.g. gas turbulence and gravitational scattering off massive gas clumps, in establishing the observed stellar velocity dispersions and orbital eccentricities. We find that models with gas fractions $>20$ per cent feature a turbulent and clumpy interstellar medium (ISM), leading to zero-age stellar velocity dispersions $\sim 20-30~{\rm km\, s}^{-1}$ and high mean orbital eccentricities. Low eccentricities cannot arise from these physical conditions. For gas fractions below $20$ per cent, the ISM becomes less turbulent, with stellar velocity dispersions $<10~{\rm km\, s}^{-1}$, and nearly circular orbits for young stars. The turbulence present in gas-rich high redshift galaxies hence acts as a `barrier' against the formation of thin discs. We compare our findings to the Milky Way's age-velocity dispersion relation and argue that velocity dispersions imprinted already at star formation by the ISM contribute significantly at all times. Finally, we show that observed orbital eccentricities in the Milky Way's thick and thin discs can be explained entirely as imprints by the star-forming ISM, rather than by mergers or secular processes.
\end{abstract}



\section{Introduction}
\label{sect:introduction}
The Milky Way's past evolution can only be studied through galactic archaeology \citep[][]{Bland:2016aa}. Using astrometric, spectroscopic and IFU surveys, the detailed kinematics and evolution of the Milky Way, and local disc galaxies in general, can now be studied in unprecedented detail \citep[e.g. Gaia, APOGEE, 4MOST, MaNGA, SAMI, ][]{Gaia2016,Apogee2017,4MOST2019,Bundy2015,Bryant2015}. At higher redshift, large galaxy surveys provide insight on the kinematic state of gas and stars in young, intensively star-forming galaxies \citep[see review by][]{Glazebrook2013}. Yet, how high redshift galaxies evolve into the well ordered spirals we observe in the local Universe is a central problem in astrophysics \citep[e.g.][]{Renaud:2021aa}. 

The Milky Way has a thin and thick disc component, with the thick disc thought to carry signatures of its high redshift formation. It is not only geometrically thicker \citep[][]{Reddy2003} and   more radially compact than the thin disc,  but also richer in $\alpha$-elements \citep[][]{Bensby2011,Bensby2014, Hayden2015}, and contains older stars \citep[$\sim 8-12$ Gyr,][]{Bensby2014,Feuillet2019}. Stellar velocity dispersions are observed to monotonically increase with the age of stellar populations in the Milky Way, with the thick disc observed to have velocity dispersions $\gtrsim 30 \kms$ \citep[][]{Mackereth:2019aa}, compared to the $\sim 10-20\kms$ for the thin disc \citep[][]{Nordstrom2004, Hayden2020}. This age-velocity dispersion relation (AVR) can arise from dynamical heating of stars, at least to ages of $\sim 8$ Gyr \citep[][]{Gustafsson2016,Yu:2018aa,Mackereth:2019aa, Ting:2018aa}. However, constant orbital heating is thought to be insufficient to explain the relatively high velocity dispersions of the thick disc \citep[e.g.][]{Sellwood2014,Mackereth:2019aa, Aumer:2017aa}. As such, a number of mechanisms for creating thick discs have been proposed in the literature, including vertical disc heating by satellite encounters \citep[][]{Quinn1993,Kazantzidis2009}, accretion of satellite stars \citep[][]{Abadi03a,Read2008}, star formation in turbulent gas rich discs \citep[][]{Bournaud09,Martig:2014aa}, bursty star formation at high redshift \citep[][]{Yu2021}, and secular formation by radial migration of kinematically hot stars from the inner to the outer disc \citep[][]{SellwoodBinney2002,Loebman2011}.   

\citet[][]{Sales:2009aa} demonstrated, using numerical simulations, how several of the above listed thick disc formation scenarios result in different distributions of stellar orbital eccentricities. As such, orbital eccentricities provide complementary information to the AVR, and can be used to constrain the scenarios. \citet{li:2018aa} used Gaia and APOGEE data to show that the thick disc is characterised by a higher fraction of stars on eccentric orbits in comparison to the thin disc. \citet{Silva:2021aa} combined this data with stellar ages and found an overlap in terms of eccentricities for old ( > 10 Gyr) stars in the thin and thick discs. This is suggests a co-formation scenario of the discs, in contrast to purely sequential formation.

The degree to which the observed thick disc properties is an imprint of the physical conditions of the star-forming ISM they are born, and hence originates from a purely in-situ process, is a debated topic. Observations of disc galaxies out to $z\sim 3$ reveals that the velocity dispersion of the ISM increases with redshift \citep[e.g.][]{Gnerucci:2011aa, Kassin2012, Wisnioski:2015aa, Mieda:2016aa}, and thick disc stellar kinematics can, in principle, be accounted for if the ISM kinematics were directly mapped onto the stars.

In this work we aim to pinpoint the role played by the turbulent and clumpy ISM in gas rich high redshift galaxies in establishing thick disc kinematics. While, in principle, cosmological simulations of galaxy formation are adequate tools to study this scenario  \citep[see e.g.][]{Brook:2004aa, Martig:2014aa,Agertz2021, Renaud:2021aa, Yu2021}, such simulations suffer from the lack of control, as multiple physical processes are simultaneously active. We overcome this problem by carrying out several non-cosmological simulations of Milky Way-mass disc galaxies. The key parameter here is the gas fraction of the galaxies, which affects the nature of gravitational instability and turbulent state of the disc \citep[][]{romeo92,Romeo2014, Renaud2021}. This approach allows us to self-consistently model the development of a turbulent ISM from disc instabilities and stellar feedback, while excluding effects such as mergers and gas accretion.

This paper is organized as follows. In Section~\ref{method} we describe the numerical method, including our choice of galaxy formation physics and the simulation setup. In Section~\ref{Results} we present the simulations results, with a focus on the dependence of stellar kinematics on the gas fraction and how our results connect to the observed properties of the Milky Way's thin and thick discs. Finally we discuss our results in Sections~\ref{discussion} and conclude in \ref{conclusion}.

\begin{table*}
	\centering
	\caption{Properties of the different simulations, all measured at 300 Myr unless otherwise indicated. Simulation names are indicated in column 1. Column 2:  gas fraction at 300 Myr; Column 3: star formation rates with the error margins representing the spread in values between 250 and 300 Myr; Column 4: mean velocity dispersion, $\pm 1$ standard deviation, of stars younger than 10 Myr; Column 5: approximate redshifts in comparison to Milky Way-like galaxies found by \citet{Dokkum:2013aa}; Column 6: approximate redshifts to allow for  birth dispersion to explain the Milky Way's age-velocity dispersion relation (see Section~\ref{sect:MWav}); Column 7: mean eccentricities of stars,  $\pm 1$ standard deviation. }
\begin{tabular}{ccccccc}
\hline
Name & $f_{\rm gas}$ (at 300 Myr) & SFR                   & $\sigma_{\rm z, birth}$ &  Redshift  & Redshift  &  Eccentricity \\
     &    per cent  & [M$_\odot$ yr$^{-1}$] & [km s$^{-1}$] &  (SFR matching)  & (AVR matching)           &                     \\ \hline
fg10 & 9   & 1 $\pm$ 0.25          & 5.2  $\pm$  0.4                   & 0.0 - 0.2  & 0.0 - 0.2  & 0.10  $\pm$ 0.06            \\
fg15 & 16 & 5 $\pm$ 0.8           & 8.6    $\pm$  1.8                & 0.5 - 1.7 & 0.1 - 0.4 &  0.33  $\pm$ 0.11          \\
fg20 & 21 & 15 $\pm$ 2            & 17.9    $\pm$ 3.5                  & 1.4 - 2.6 & 0.4 - 0.8 &  0.49   $\pm$ 0.16             \\
fg30 & 29 &25 $\pm$ 5            & 27.3      $\pm$ 3.8                   & 1.8 - 2.6 & 0.8 - 2 &  0.53   $\pm$ 0.21              \\
fg45 & 46 &  50 $\pm$ 10           & 25.8   $\pm$ 4.3                & 1.8 - 2.6 & 0.8 - 2 &0.47      $\pm$ 0.22         \\ \hline
\end{tabular}
\label{tab:sfr}
\end{table*}

\begin{figure*}
\centering
\setlength\tabcolsep{2pt}
\includegraphics[ trim={0cm 0cm 0cm 0cm}, clip, width=1\textwidth, keepaspectratio]{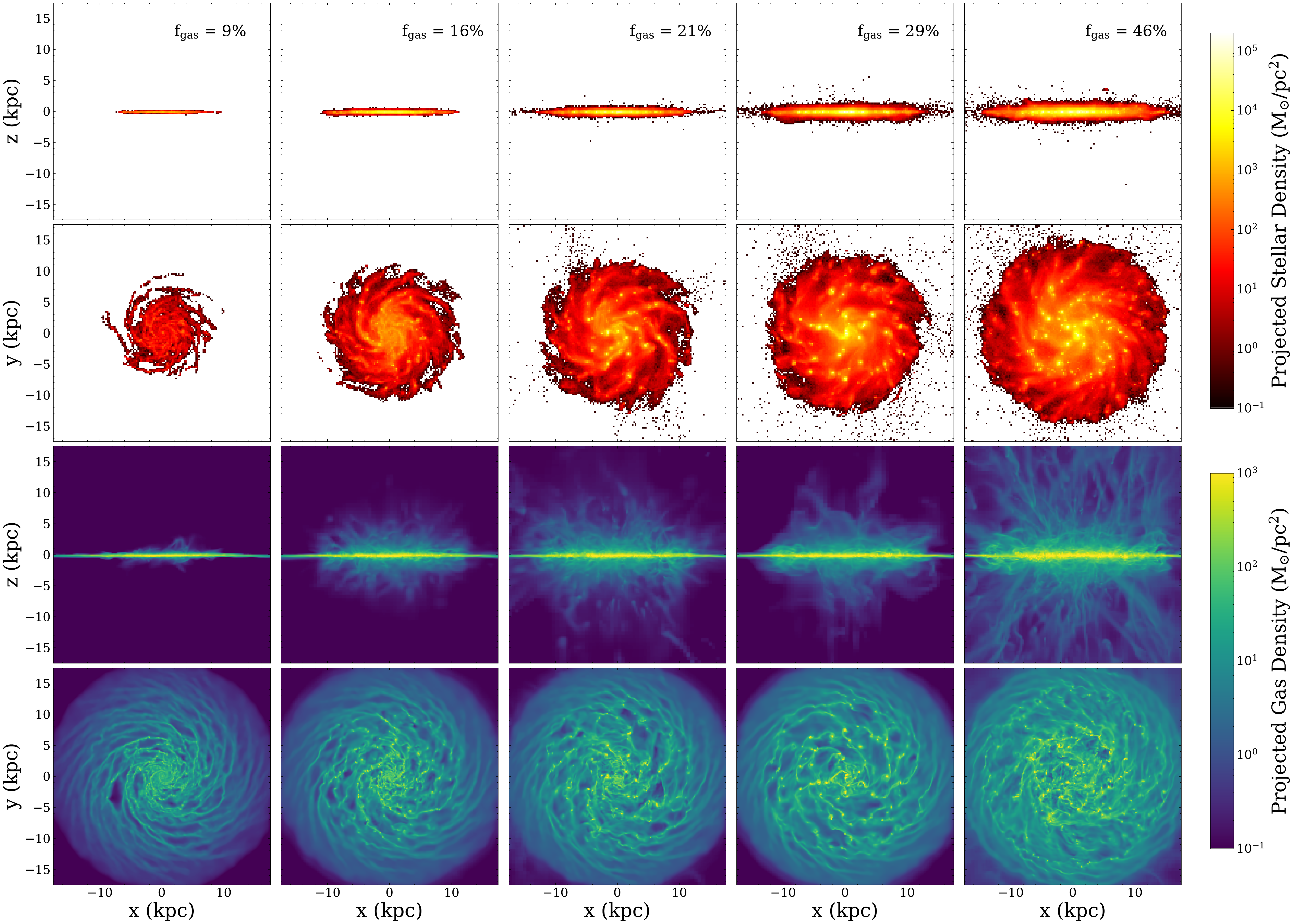}
\caption{Stellar (upper) and  gas (under) density projections of the simulated galaxies at 300 Myr. The initial gas fraction increases, from left to right, from 10 to 70 per cent. The gas density projections show the presence of flocculent spiral structure in all galaxies. In the higher gas fraction regimes, there is a clear presence of massive clumps and more irregular ISM. } \label{fig:map}
\end{figure*}

\section{Method}
\label{method}
We use the hydrodynamic+$N$-body code {\small RAMSES} \citep{teyssier02} to model Milky Way-mass galaxies in isolation.  The fluid dynamics of the baryons is calculated using a second-order unsplit Godunov method, with the equation of state of an ideal mono-atomic gas with an adiabatic index $\gamma=5/3$. The code accounts for metallicity dependent cooling by using the tabulated cooling functions of \citet{sutherlanddopita93} for gas temperatures $>10^4$ K, and rates from \citet{rosenbregman95} for cooling down to lower temperatures. The collisionless dynamics of stellar and dark matter particles is evolved using the particle-mesh technique \citep{Hockney1981}, with gravitational accelerations computed from the gravitational potential on the mesh using a multi-grid method \citep{GuilletTeyssier2011}. 

We achieve high resolution in high density regions using adaptive mesh refinement. A cell is split if its baryonic mass (gas and stars) exceeds 4014 $\Msol$. In addition, a cell is allowed to refine if it contains more than 8 dark matter particles. This allows the local force softening to closely match the local mean inter-particle separation, which suppresses discreteness effects \citep[e.g.][]{Romeo08}. The maximum refinement level is set to allow for a mean  physical resolution of $\sim 9 \pc$ in dense gas.

The adopted star formation and feedback physics is presented in \citet[][see also \citealt{AgertzRomeoGrisdale2015, Agertz2020}]{Agertz2013}  Briefly, star formation is treated as a Poisson process, sampled using star particles of $1000 \Msol $, occurring on a cell-by-cell basis according to the star formation law:
\begin{equation}
	\dot{\rho}_{\star}= \epsilon_{\rm ff}\frac{ \rho_{\rm g}}{t_{\rm ff}} \quad {\mbox{for}} \quad  \rho>\rho_{\rm SF}.
	\label{eq:schmidtH2}	
\end{equation}
Here $\dot{\rho}_{\star}$ is the star formation rate (SFR) density, $\rho_{\rm g}$ the gas density, $t_{\rm ff}=\sqrt{3\upi/32G\rho_{\rm g}}$ is the local free-fall time of the gas, $\rho_{\rm SF}=100~{\rm cm}^{-3}$ is the star formation threshold, and $\epsilon_{\rm ff}$ is the local star formation efficiency per free-fall time of gas in the cell. Observationally, $\epsilon_{\rm ff}$ averages at about $1$ per cent in Milky Way giant molecular clouds (GMCs \citealt[][]{krumholztan07}), albeit with a spread of several dex \citep[][]{Murray2011b,Lee2016}. \citet{Grisdale2019} demonstrated how high efficiencies ($\epsilon_{\rm ff}\sim 10$ per cent) on scales of parsecs provides a close match to  observed efficiencies on scales of individual GMCs, as well as for the properties of the cold ISM \citep{Grisdale2017,Grisdale2018}. Motivated by these findings, we adopt $\epsilon_{\rm ff}=10$ per cent.  
Each formed star particle is treated as a single-age stellar population with a \citet{chabrier03} initial mass function. We account for injection of energy, momentum, mass, and heavy elements over time from core-collapse supernovae (SNe) and SNIa, stellar winds, and radiation pressure on the surrounding gas. Each mechanism depends on the stellar age, the mass and stellar metallicity (through the metallicity dependent age-mass relation of \citealt{Raiteri1996}, calibrated on the stellar evolution code {\small STARBURST99}, \citealt{Leitherer1999}). 

SNe explosions are modelled as discrete events, and we follow \citet[see also \citealt{Martizzi2015}]{KimOstriker2015} in injecting the  momentum generated during the Sedov-Taylor phase if the cooling radius\footnote{The adopted definition of the cooling radius, for a supernova explosion with energy $E_{\rm SN}=10^{51}$ erg, is $r_{\rm cool}= 30 n_0^{-0.43}(Z/Z_\odot+0.01)^{-0.18}$ pc  \citep[e.g.][]{Cioffi1988,Thornton1998,KimOstriker2014}}  is not resolved by at least 6 grid cells. Otherwise, we inject $10^{51}$ ergs of thermal energy \citep[see][for details]{AgertzRomeoGrisdale2015} and allow for the hydrodynamic solver to track the buildup of momentum. 

\subsection{Galaxy simulations} \label{ramses}
The simulations have galactic parameters chosen to mimic a present-day Milky Way-mass disc galaxy. The simulations are carried out in isolation, with the simulated galaxy positioned at the center of box with sides 600 kpc. While this means we are neglecting cosmological effects such as mergers and inflows, it has the benefit of allowing for high numerical resolution, as well as being a controlled experiment. The initial conditions are based on the isolated disc galaxy in the AGORA project \citep{Kim:2014aa, Kim:2016aa}, following the methods described in \citet{Hernquist:1993aa} and \citet{Springel:2000aa}. A Navarro-Frenk-White profile \citep{Navarro:1996aa} is adopted for the dark matter halo, with a concentration parameter  $c=10$, and a virial mass\footnote{These choices yield a virial circular velocity v$_{200}$ = 150 km/s and virial radius r$_{200}$ = 205 kpc.} of M$_{200} = 1.1 \times 10^{12}$ M$_\odot$. 

The initial stellar and gaseous components follow exponential surface density profiles with scale lengths $r_{\rm d} = 3.4$ kpc and scale heights $z_{\rm d} = 0.1r_{\rm d}$. The gas disc is initialised at a temperature of $T=10^4~$K. The total disc mass is $4.5 \times 10^{10}~{\rm M_\odot}$, with the fraction of mass in the gas disc varying between simulations. This allows us to isolate the effect of the gas fraction, which changes the nature of gravitational instability and ISM turbulence on the resulting stellar kinematics (see \citealt{Renaud2021}, hereafter \citetalias{Renaud2021}). This approach and the numerical recipe is the same as in  \citetalias{Renaud2021}, but we use a different, larger set of initial conditions. The gas fraction is defined as

\begin{equation}
\label{eq:fgas}
f_{\rm gas}=\frac{M_{\rm gas}}{M_{\star}+M_{\rm gas}}, 
\end{equation}
where $M_{\star}$ and $M_{\rm gas}$ is the total stellar and gas mass in the disc and the bulge, respectively. The gas fraction here acts as a proxy for cosmological epoch, with initial disc gas fractions of 10, 20, 30, 50 and 70 per cent. The bulge-to-disc mass ratio is 0.125 and the bulge mass profile is that of \citet{Hernquist:1990aa} with scale-length 0.1$r_{\rm d}$. The halo and stellar disc are represented by 10$^6$ particles each, and the bulge consists of 10$^5$ particles.

The galaxies were simulated for 300 Myr, with the initial $\sim 100$ Myr evolved at lower spatial resolution and without stellar feedback. This allows for a gentle transition from the initial conditions into a new equilibrium (see also Ejdetj\"arn et al. in prep.). All the analysis is carried out at the end of the simulation by only considering the stars formed after this relaxation period. Given the choice of initial conditions, we emphasise that our study is relevant for understanding the origin of stellar kinematics in typical $L^\star$ disc galaxies, and useful for a comparison to Milky Way-mass disc galaxies an their progenitors. 

We assign approximate cosmological epochs to our simulated galaxies by comparing their star formation rates\footnote{Computed as the average SFR between 250-300 Myr after simulation start.} (SFRs) to observationally inferred star formation histories of Milky Way-mass galaxy progenitors by \cite{Dokkum:2013aa}. Dokkum et al. used deep photometric catalogs from the 3D-HST and CANDELS Treasury surveys up to $z\sim 2.3$ to determine SFRs for galaxy populations that match the abundance of present-day Milky Way-mass galaxies. Each simulation hence has a corresponding redshift range which we summarise, together with other galaxy properties, in Table \ref{tab:sfr}.

\begin{figure}
\centering
\setlength\tabcolsep{2pt}%
\includegraphics[ trim={0cm 0cm 0cm 0cm}, clip, width=0.48\textwidth, keepaspectratio]{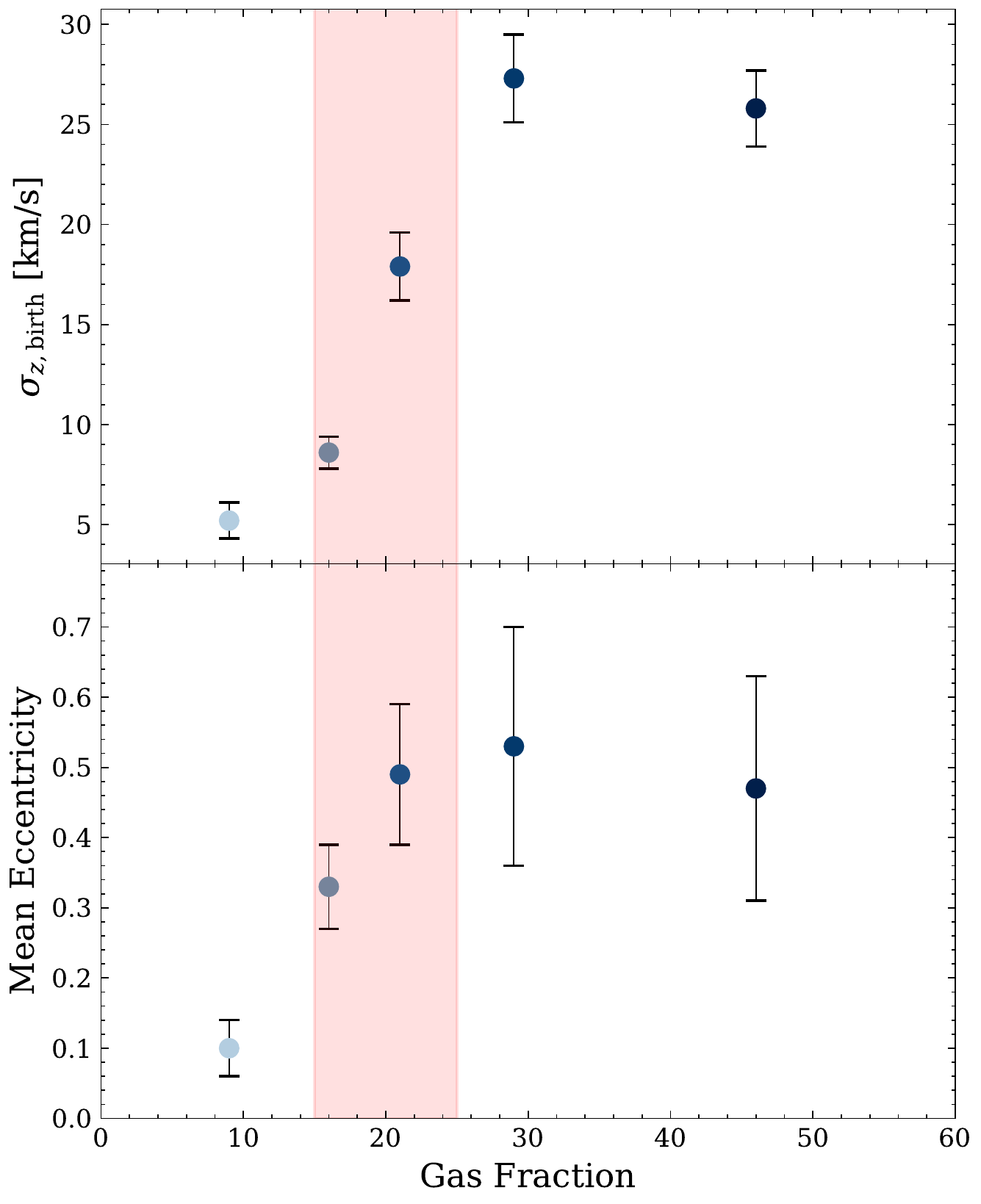} 
\caption{The mean birth velocity dispersion (upper panel) and  eccentricity (lower panel) as a function of the galaxy gas fraction. The vertical error bars correspond to the standard deviation of the measurements. The horizontal error bars display the range of gas fractions each simulation went through. The colored bar indicates the transition in kinematic properties around $f_{\rm gas} \approx 20$ per cent \citep[recently highlighted by][]{Renaud2021}.} 
\label{fig:eccoz}
\end{figure}

\section{Results} 
\label{Results}
We begin the analysis by studying stellar and gas density maps of the simulations at 300 Myr, shown in Figure \ref{fig:map}. At this point in time, star formation and feedback have decreased the initial gas fractions (computed in a slab covering $\pm 2.5\kpc$ above and below the mid-plane of the disc\footnote{Almost identical values for the gas fractions are found when considering the cold gas only ($T\leq 10^4$ K).}) to $9,16, 21, 29$ and $46$ per cent. We henceforth use this evolved gas fractions to label our simulations: fg10, fg15, fg20, fg30, and fg45 respectively (see Table \ref{tab:sfr}).

With increasing gas fraction, a larger portion of the ISM reaches high, star forming densities (see \citetalias{Renaud2021}). In the highest gas fraction galaxy, the SFR settles to $\sim 50~$ M$_\odot$ yr$^{-1}$, in contrast to $\sim 1~$ M$_\odot$ yr$^{-1}$ in the most gas poor case (see Table~\ref{tab:sfr}). The resulting increase in stellar feedback activity in the gas rich galaxies produces vigorous galactic outflows, as is evident from the edge-on views of the gas density fields in Figure \ref{fig:map}. Together with gravitational instabilities, feedback drives ISM turbulence, with higher velocity dispersions measured in the cold star forming ISM in the gas rich galaxies ($\sim 30\kms$ in fg30) compared to the gas poor ones ($<10\kms$ in fg10, see Ejdetj\"arn et al. in prep. for a detailed analysis). 

For $f_{\rm gas}\lesssim 20$ per cent, a spiral structure characterizes the galaxy morphology, while massive clumps dominate the morphology at higher gas fractions, in line with observations of gas-rich high redshift galaxies \citep[see e.g.][]{Elmegreen:2009aa, Dessauges:2019aa}. This transition in terms of morphology around a gas fraction of $\sim$ 20 per cent, for a disc galaxy of this size and mass, is expected from the evolution of the regimes of gravitational instability (see \citetalias{Renaud2021}), and leaves distinct imprints in stellar kinematics, which we turn to next.

\subsection{Imprints on stellar kinematics} 
\label{imprints}
The upper panel of Figure \ref{fig:eccoz} shows the mean vertical stellar velocity dispersion of all stars in the disc as a function of gas fraction, at the same simulation time as in Figure~\ref{fig:map} ($300$ Myr). Only stars younger than 10 Myr are considered, which allows us to quantify the imprint on stellar kinematics at star formation, rather than secular heating.

As seen in the figure, stars born in higher gas fraction environments have a significantly higher vertical velocity dispersion at birth, with $\sigma_{z, {\rm birth}}$ plateauing at $\sim 25 - 30\kms$ for $f_{\rm gas}\gtrsim 30$ per cent. As discussed above, a gas fraction of 20 per cent marks a transition (indicated by the vertical red band), with lower gas fractions leading to a less turbulent, more ordered state in the disc, with $\sigma_{z, {\rm birth}}\lesssim 10\kms$ for $f_{\rm gas}\lesssim 20$ per cent. 

\begin{figure}
\centering
\setlength\tabcolsep{2pt}%
\includegraphics[ trim={0cm 0cm 0cm 0cm}, clip, width=0.48\textwidth, keepaspectratio]{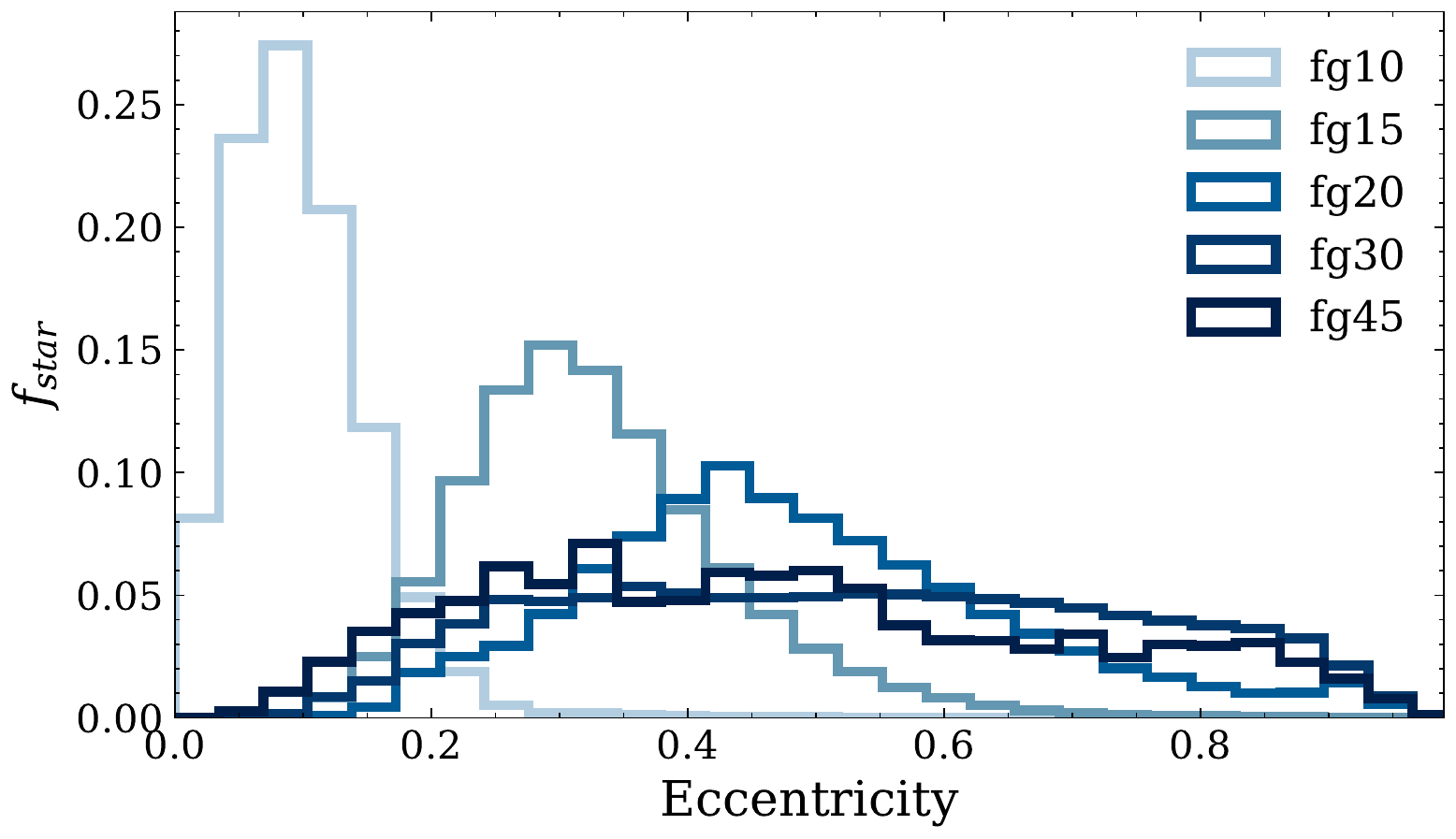}
\caption{The full eccentricity distributions  for the different simulations, all stars are included.  }
\label{fig:eccdist}
\end{figure}

The vertical velocity dispersion is not the only quantity that carries a distinct imprint of the ISM conditions. The bottom panel of Figure \ref{fig:eccoz} shows the mean eccentricity\footnote{For each star, the eccentricity is computed as $e=\frac{r_{\rm a}-r_{\rm p}}{r_{\rm a}+r_{\rm p}}$ , with $r_{\rm a}$ and  $r_{\rm p}$ being the apocenter and pericenter of an individual orbit, respectively.} of stellar orbits as a function of the gas fraction. The vertical error bars indicate the standard deviation of the full eccentricity distributions shown in Figure~\ref{fig:eccdist}. Low gas fractions entail low eccentricities, with fg10 featuring, on average, close to circular orbits with mean stellar eccentricities $\sim 0.1$. 

As the gas fraction is increased, a similar transition as previously discussed occurs; the eccentricity distribution shifts to higher values, plateauing at an average of $\sim 0.5$ for all simulations with $f_{\rm gas}\gtrsim 20$ per cent. In these gas rich galaxies, the eccentricity distributions are also significantly broader and flatter compared to the low gas fraction counterparts (see Figure~\ref{fig:eccdist}), with an almost equal likelihhood to find stars with eccentricities anywhere in between 0.2 and 0.8 for $f_{\rm gas} > 20$ per cent. The origin of this marked shift towards higher eccentricities is likely the increase in number, and masses, of dense clumps that  gravitationally scatter newly born stars onto eccentric orbits. However, a more detailed analysis, tracing individual stars back to their birth environments, is necessary to more conclusively characterise this process, which we leave for future work. Without a significant change to the galactic potential or violent scattering events in the later life of the galaxy, such as galaxy mergers, these orbital properties will remain to be observed today as imprints of the gas-rich, high-redshift Universe.

\subsection{Implications for the Milky Way} \label{MWimp}
\label{sect:MW}
\subsubsection{The age-velocity dispersion relation}
\label{sect:MWav}
As discussed in Section~\ref{sect:introduction}, recent astrometric and spectroscopic surveys such as Gaia \citep[][]{Gaia2016} and APOGEE \citep[][]{Apogee2017} have allowed for accurate positions and 3D velocities of millions of stars in the Milky Way. \citet{Mackereth:2019aa} used this information, together with estimated stellar ages (using a Bayesian neural network model trained on asteroseismic ages) to construct an AVR for the thick and thin discs.

\begin{figure}
\centering
\setlength\tabcolsep{2pt}%
\includegraphics[ trim={0cm 0cm 0cm 0cm}, clip, width=0.48\textwidth, keepaspectratio]{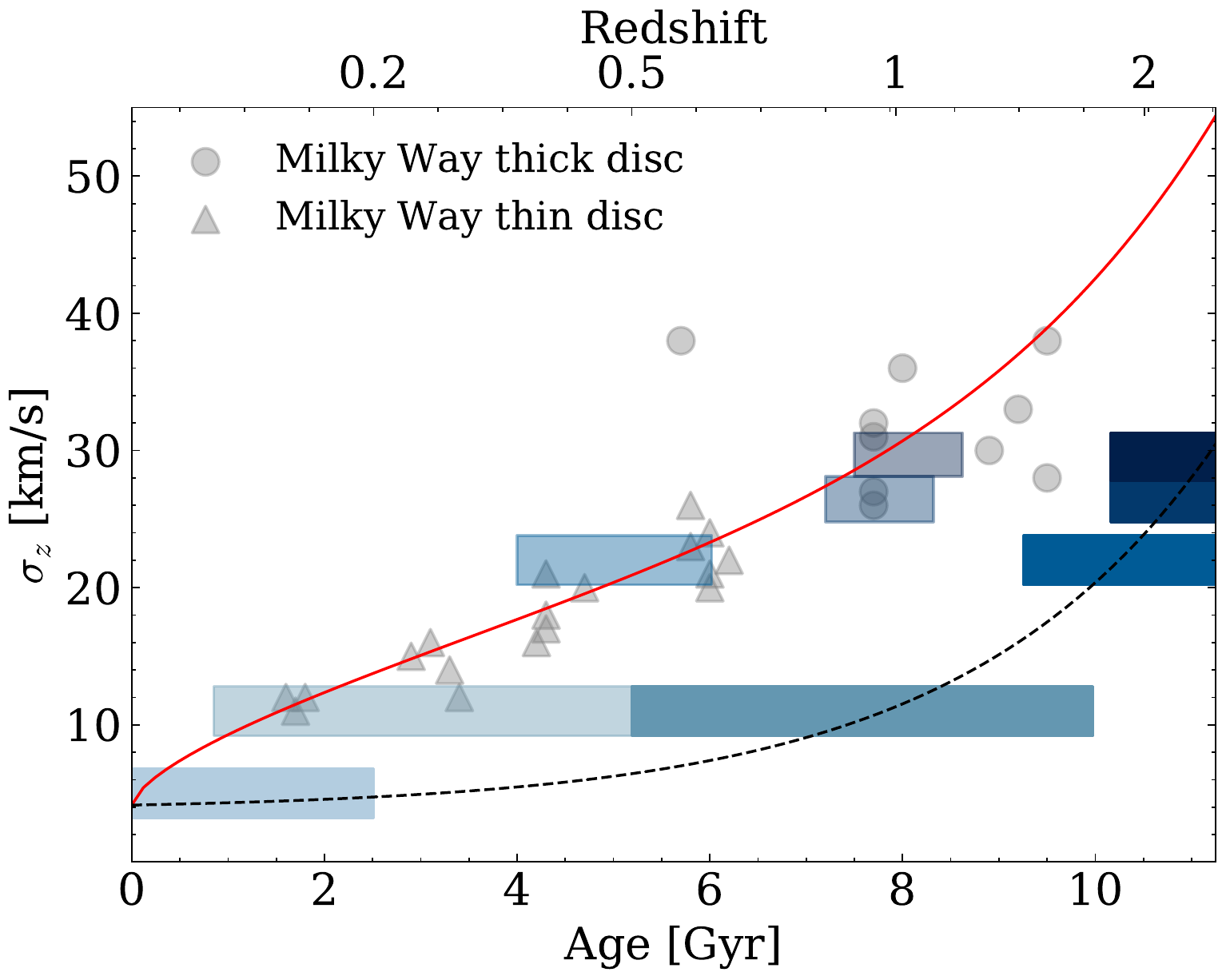}
\caption{The Milky Way's AVR for the thin and thick disc \citep[gray symbols,][]{Mackereth:2019aa}, together with birth velocity dispersions from the simulations, where the associated cosmic epochs are chosen to match the Milky Way data (semi-transparent boxes). The non-transparent boxes show the simulated data at epochs obtained by matching SFRs to observations of Milky Way progenitors, see main text for details. The black dashed line is an exponential fit to this data, and by adding secular heating to this relation, an AVR is obtained (red line) that jointly matches the Milky Way's thick and thin discs.} 

\label{fig:moving}
\end{figure}

The symbols in Figure \ref{fig:moving} show the AVR from Mackereth et al. for stars in the Solar neighbourhood (i.e. galactocentric radii 7-9 kpc). Our simulations are shown as shaded regions with the dashed line being a non-linear least squares fit of the form $\sigma_{z,{\rm birth}}\propto \exp( \rm {age} )$. The horizontal ranges of the simulation data refer to the approximate redshifts obtained from matching the simulated SFRs to the observed SFR-redshift relation for Milky Way-mass progenitors (see Table~\ref{tab:sfr}). We note that this is a qualitative comparison; we do not account for any galactocentric radial dependencies in the AVR that exist in the observational or simulation data, nor do we attempt to account for radial migration \citep[][]{SchonrichBinney2009} that could shape the AVR we observe today \citep[but see][]{Minchev2012,Agertz2021}. 

As seen in Figure \ref{fig:moving}, stellar velocity dispersions imprinted at birth are not sufficient to account for the observed trend. While this discrepancy can be mitigated by accounting for the large observational age uncertainties for old stars\footnote{While uncertainties on stars younger than 5 Gyr are on the order of $\sim 30$ per cent in the method of \citet{Mackereth:2019aa}, estimated ages of stars with true ages $> 10$ Gyr can be under-estimated by as much as 3.5 Gyr. Such large uncertainties can mitigate the discrepancies, at least at $z>1$, between our simulations and the observations.}, we explore two alternative ways of explaining the difference between the observed and simulated AVR. While both options can contribute to the observed velocity dispersions, we consider them one by one independently for simplicity.

\begin{enumerate}
\item The method for associating redshifts to the simulations is incorrect, and higher gas fractions were present at recent times. The corresponding redshifts for matching the observed AVR with birth dispersions {\it alone} are shown as semi-transparent boxes in Figure \ref{fig:moving} (see also Table~\ref{tab:sfr}).
 
\item Additional secular heating plays a key role in shaping the AVR. 
\end{enumerate}

The first option cannot be ruled out. In fact, by allowing for $f_{\rm gas}\approx 20$ and $\approx 30$ per cent at $z\sim 0.5$ and $1$, respectively, hence matching the AVR, our results are in closer agreement with the cold gas reservoirs observed in high redshift galaxies \citep[see a compilation of molecular gas fraction over time by][]{Saintonge:2013aa}. It is therefore possible to explain the observed AVR of the Milky Way by solely invoking birth velocity dispersions and an evolving galactic gas fraction. However, we note that the evolution of the gas fraction in galaxies depends on a large number of factors, including the stellar mass and environment \citep[e.g. ][]{Santini2014, Schreiber2015, Genzel2015,Tacconi2018}, as well as observational selection biases. To relate the specific gas fraction evolution of the Milky Way to observed (average) trends is therefore non-trivial, and beyond the scope of this work.
 
If we instead retain the originally assigned cosmic epochs to the simulations, secular heating via gravitational scattering of stars off GMCs and massive clumps, spiral arms, and bars occurring over many billions of years, must contribute to the AVR \citep[see review by][]{Sellwood2014}. \citet[][]{Aumer:2017aa} used  $N$-body simulations of galactic discs to study secular heating, and demonstrated that their simulated AVRs are well fitted by a relation that combines secular and birth velocity dispersions, of the form

\begin{equation}
\label{eq:sigmaanalytical}
\sigma(t) = \sigma_{0}\Big( \frac{{\rm age}}{10 \text{ Gyr}} \Big)^{\beta} + \sigma_{z, {\rm birth}}(t) [{\rm km/s}].
\end{equation}
Here $\sigma_{0}$ scales the secular heating term, $t$ is the age of the stars, the power law index $\beta$ describes the efficiency of vertical heating, and $\sigma_{z, {\rm birth}}(t)$ is the birth velocity dispersion of stars. The last term is often assumed to be a constant, e.g. $6\kms$ \citep[][]{Aumer:2016ab}. However, our results show that this underestimates the contribution of birth velocity dispersions in the Milky Way's more gas rich past.

The red line in Figure \ref{fig:moving} shows the  total heating using Equation~\ref{eq:sigmaanalytical} with  $\sigma_{0}=22.2 \kms$, $\beta=0.65$ \citep[compatible with the results from][]{Aumer:2016ab} and $\sigma_{z, {\rm birth}}(t)$ obtained from our simulations (dashed line). This AVR provides a good match to the Milky Way data, and jointly reproduces the thick and thin disc data points in the Solar neighbourhood. This joint match cannot be achieved without the contribution from a varying $\sigma_{z, {\rm birth}}(t)$ in Eq. \ref{eq:sigmaanalytical}.

We note that outliers exists to the mean relation, e.g. 6 Gyr old thick disc stars. Such populations could be a signature of co-eval stars originating from different birth environments (e.g. inner vs. outer disc, with different [$\alpha$/Fe]) in terms of gas fractions, or regions with different strengths of secular heating \citep[][]{Aumer:2016aa, Agertz2021}, that subsequently mix radially into the Solar neighbourhood \citep[e.g.][]{Minchev2013,Frankel2018,Minchev2018,Mikkola2020}. 

In summary, we find that birth velocity dispersions alone can explain the Milky Way's AVR. The AVR is then intimately linked to the Milky Way's gas fraction evolution, and the associated level of ISM turbulence and fragmentation physics. On the other hand, if the Milky Way's gas fraction was lower at early times, as suggested from matching simulated SFRs to observations of Milky Way mass progenitors, a combination of evolving birth velocity dispersions and secular processes must be invoked to explain the scaling and shape of the AVR. However, the birth velocity dispersion is even in this case always a crucial component, being responsible for at least 30 per cent of the total heating at all epochs. Furthermore, it accounts for at least 50 per cent of the vertical kinematics for stars belonging to the old ($\gtrsim 9$ Gyr) thick disc, as well as stars younger than $\sim 1$ Gyr as secular heating has not yet had time to significantly affect their kinematics. 

\begin{figure}
\centering
\setlength\tabcolsep{2pt}%
\includegraphics[ trim={0cm 0cm 0cm 0cm}, clip, width=0.48\textwidth, keepaspectratio]{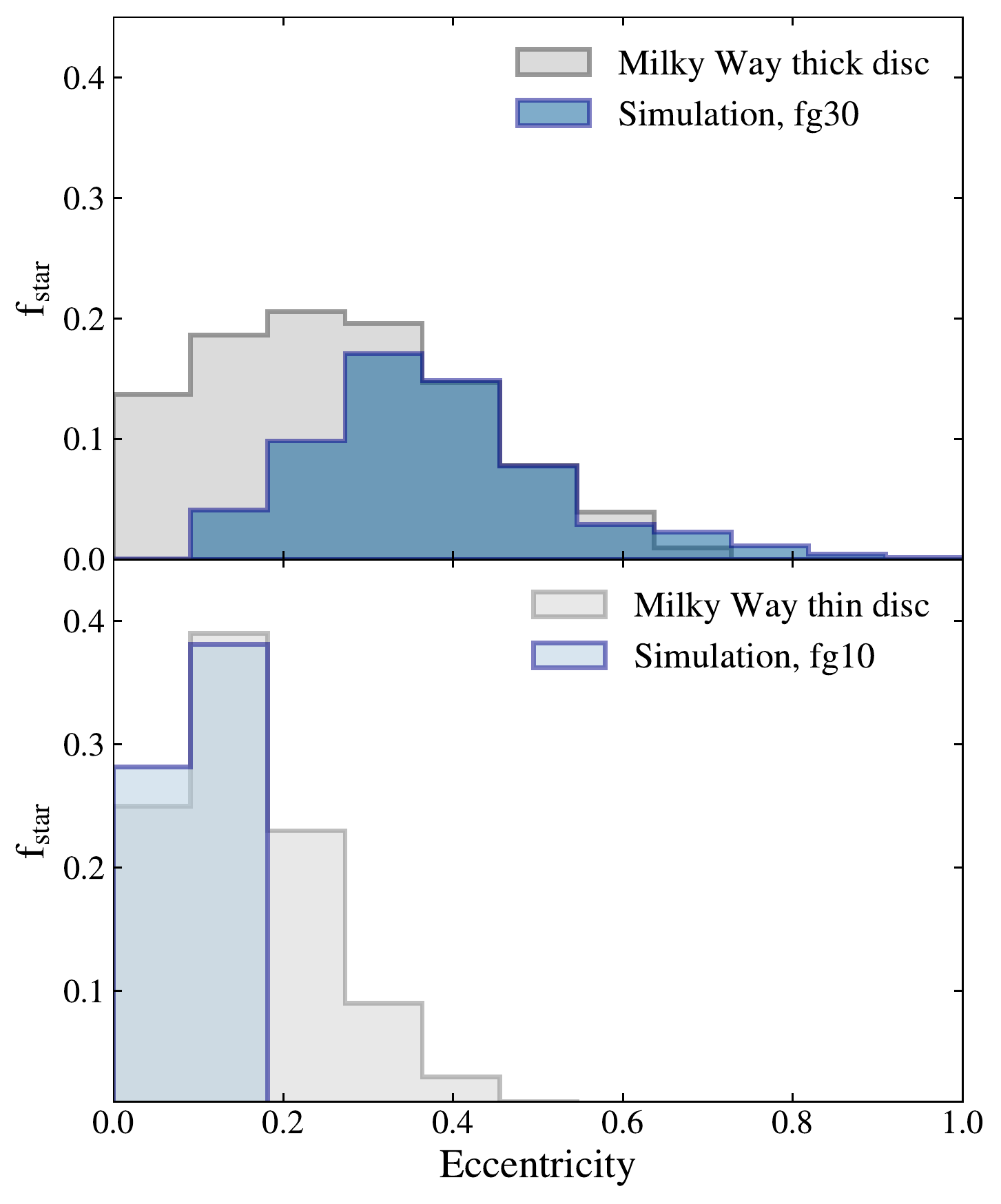}
\caption{Normalised distributions of Orbital eccentricity distributions from \citet{li:2018aa} in gray for the thick (top) and thin (bottom) discs. Coloured distributions are from two of our simulations, fg30 and fg10 respectively. The gas rich fg30 simulation can account for the entire high eccentricity tail of the Milky Way's thick disc distribution but shows a deficit of low eccentricity populations.}
\label{fig:bins}
\end{figure}

\subsubsection{Formation environments of the thin and thick discs constrained by eccentricities}
\citet{li:2018aa} used Gaia and APOGEE data to study eccentricities of stars in the Milky Way's thin and thick discs. The discs were defined by their [$\alpha$/Fe] chemical abundances, analogous to \citet{Mackereth:2019aa}. Figure \ref{fig:bins} shows the eccentricity distributions from \citet{li:2018aa} for the thick (top) and thin (bottom) disc. The thick disc is characterised by a higher fraction of stars on  eccentric orbits in comparison to the thin disc. 

The sensitivity of orbital eccentricities to the gas fraction allows us to to constrain the formation environments of the discs. However, we refrain from directly fitting the distributions obtained from our simulation suite to the observations; the sensitivity to adopted galactocentric radial cuts and unaccounted processes, such as radial migration over a long period, makes such an approach degenerate. Instead, we overlay the eccentricity distribution from fg30 (`gas rich disc') and fg10 (`gas poor disc') to perform a more qualitative comparison. Only stars present at galactocentric radii $7-9~\kpc$ in the simulations are considered.

As seen in Figure~\ref{fig:bins}, the tail of high eccentricities ($\gtrsim 0.4$) in the thick disc distribution is well matched by stars that have formed in the gas rich disc. This means that cosmological processes, such as mergers, are not necessary to explain such high eccentricities. We re-emphasize that this is not due to slow secular processes \citep[e.g. radial migration,][]{Sales:2009aa}, but an imprint of the turbulent and clumpy ISM. However, to explain the existence of eccentricities $\lesssim 0.1-0.2$, lower gas fractions are required. This may indicate that the thick disc's gas fraction decreased rapidly in the early Universe, while retaining a high [$\alpha$/Fe]. The redshifts obtained from our SFR matching are compatible with this notion, as $f_{\rm gas}$ decreases from 50 to 20 per cent in less than a Gyr at $z\sim 1.5$ (see Table~\ref{tab:sfr}). Alternatively, as discussed in Section~\ref{sect:MWav}, different gas fractions may have been present simultaneously in different regions of the galaxy when the [$\alpha$/Fe]-enhanced thick disc formed. 

The thin disc distribution is well matched by the gas poor disc for eccentricities $<0.2$, but higher gas fractions are required to fit the entire distribution. However, it is unlikely for the thin disc to have formed any significant fraction of its stars when $f_{\rm gas}\gtrsim 20$ per cent: such environments produce too many stars with higher than observed eccentricities.

The existence of a population of low eccentricity stars, originating from low gas fraction conditions, in both the thin and thick disc, is compatible with them having overlapping formation times. Indeed, \citet{Silva:2021aa} combined stellar age estimates, orbital eccentricities and [$\alpha$/Fe] from APOGEE and Gaia, to argue for a co-eval formation scenario $9-10$ Gyr ago. At this epoch, the Milky Way's inner and outer disc must have had similar gas fractions, but different levels of $\alpha-$enhancement. Such an inner-outer disc  dichotomy, in terms of [$\alpha$/Fe], is recovered in cosmological simulations of galaxy formation around the time of the last major merger \citep[][]{Agertz2021,Renaud:2021aa,Renaud2021b}, with subsequent radial mixing bringing stars from both regions into the Solar neighbourhood. 

\section{Discussion}
\label{discussion}
The physical origin of the diversity of orbital and kinematical properties observed in the stellar populations in the solar neighbourhood has been a long lasting debate in the literature. Models and simulations have emphasised the role of dynamical heating by a bar and/or spiral arms \citep[e.g.][]{Grand:2016aa,Aumer:2016aa}, or via radial migration \citep[][]{SchonrichBinney2009}. Cosmological simulations also point out that the transition from a merger-dominated growth of the galaxy to a more quiescent, secular evolution coincides with the transition from formation in the turbulent thick disc, subject to repeated tidal excitations to that in the dynamically colder thin disc \citep[][]{Renaud:2021aa,Bird:2021aa}. However, simulations run without the cosmological context have already pointed out comparable transitions driven by the intrinsic evolution of the regime of star formation \citep[e.g.][]{Clarke:2019aa,Khoperskov2021}. Our results complement this by showing that the shapes of the AVR and the distribution of orbital eccentricities can be explained by the intrinsic, natural, evolution of the gas content of the galaxy, without the need to invoke secular dynamical processes like migration, nor external effects like mergers. 

The decrease of the gas fraction expected along galaxy evolution changes the regime of disc instabilities and, in turn, of the properties of the ISM turbulence (\citetalias{Renaud2021}, Ejdetj\"arn et al. in prep). A similar gas fraction dependence is expected\footnote{\citet{HaywardHopkins2017} showed that for $f_{\rm gas}\gtrsim 30$ per cent, feedback is likely to be trapped in the dense ISM, leading to high levels of turbulence. At lower gas fractions, feedback ejecta is able to leak out of the galaxy, hence not injecting momentum into the ISM, with lower levels of turbulence as a result.} when stellar feedback is the dominant turbulence driver \citep{HaywardHopkins2017}. As such, the existence thin galactic discs is intimately connected to the evolution of galactic baryon fractions \citep[see e.g.][]{agertzkravtsov2016} and the way in which gas is accreted and lost in feedback driven outflows. \citet[][]{Yu2021} used cosmological simulations of Milky Way-mass galaxies to show that thin galactic discs cannot form unless the gas cooling time in the circum-galactic medium (CGM) is longer than the dynamical time \citep[see also][]{Stern2021}. The authors argued that for short CGM cooling times, gas is accreted more rapidly, leading to burstier star formation and thick disc formation. In light of our work, it is not necessarily star formation bursts that lead to thick discs, but rather the high levels of ISM turbulence and the presence of massive clumps, that in turn imprint thick disc stellar kinematics. In this picture, the state of the CGM is a crucial ingredient, as it dictates, together with the strength of galactic outflows, the ability to form and maintain gas rich turbulent discs.

Our result serve as a proof-of-concept that such an ``historical'' reconstruction can suffice to reproduce the observables. However, key aspects neglected here for simplicity would likely alter our conclusions. In particular, our simulations do not represent different snapshots along the evolution of the same galaxy, as the mass and size are unchanged, while a real galaxy would grow. A major contribution to this growth is that of galaxy mergers, which are know to stir the ISM and boost the turbulence precisely at the epochs the most fertile in star formation. By not capturing this aspect, our model underestimate the velocity dispersion of the stars born during the merger-dominated phase (e.g. $z\gtrsim 1$ for the Milky Way).

Finally, the accretion of gas along cosmological filaments, also neglected in our work, is suspected to influence the injection scale of turbulence (Agertz et al, in prep.), alter the relation between gas fraction and velocity dispersion, and thus skew the AVR. Therefore, our results should be interpreted as an illustration of the importance of the secular evolution of the gas content on the dynamics of the stellar populations, among other coeval processes.

The sharp transition seen for gas fractions $\approx 20$ per cent constitutes a strong indication of how fast this process contributes to the switch between dynamically hot to cold populations. We note that, interestingly, in the Milky Way such a transition is estimated to occur approximately at the same epoch as the last major merger, and that the two changes of regimes likely conspire in creating the observed dynamical bimodality of stellar populations.

\section{Conclusions}
\label{conclusion}
Using a series of hydrodynamical simulations of isolated, Milky Way-mass disc galaxies, we study the impact of the gas fraction on the stellar kinematics. Our main conclusions are as follows.

\begin{itemize}
    \item Newly formed stars in disc galaxies with high gas fractions feature high vertical velocity dispersions. This is a direct imprint of the turbulent ISM. Simulations with $f_{\rm gas}\gtrsim 30$ per cent generate vertical stellar velocity dispersions $\sigma_z\sim$ 30 km/s already at star formation, well above the observed kinematics in the Milky Way's thin disc, but compatible with the thick disc \citep[][]{Mackereth:2019aa}. In contrast, simulations with $f_{\rm gas}\lesssim 20$ per cent form stars with $\sigma_z\lesssim$ 10 km/s, in line with young stars in the Milky Way's thin disc. 

    \item The $\sigma_z$ imprinted by the ISM is in principle sufficient for explaining the Milky Way's age-velocity dispersion relation (AVR). However, by assigning a lookback time to the simulations, by matching their SFRs to those of observed Milky Way progenitors, additional secular heating is likely required to explain the AVR. However, the birth velocity dispersion is always a crucial component for total level of stellar heating, especially in the early Universe ($> 9-10$ Gyr in the past).
    
    \item Orbital eccentricities of young stars depend on the disc gas fraction. Stars born in galaxies with $f_{\rm gas} \sim 10$ per cent  feature stars on mostly circular orbits, with mean orbital eccentricities $\sim 0.1$. As the gas fraction is increased, stronger gravitational scattering allows for wider eccentricity distributions with higher mean values. For $f_{\rm gas}>20$ per cent, we find a `saturation effect', with similar (wide and flat) eccentricity distributions, and mean eccentricities of young stars $\sim$ 0.5. 
    
\item We find that the high eccentricity tail of the Milky Way's thick disc distribution in the Solar neighbourhood can be fully explained with star formation in a turbulent high gas fraction disc ($f_{\rm gas}> 30$ per cent). Both the thin and the thick disc require some contribution from a gas-poor regime to fully characterise their eccentricity distributions. This notion is compatible with a thin/thick disc co-formation scenario \citep[e.g.][]{Agertz2021, Renaud:2021aa,Renaud2021b, Silva:2021aa} in which a range of ISM conditions existed simultaneously, with radial migration later mixing stars across the galaxy.

\end{itemize}

We conclude that the decrease of gas fraction, as expected along galaxy evolution, changes the regime of disc instabilities, properties of the ISM turbulence, and in turn the imprinted kinematical properties of newly formed stellar populations. A key result is that this is not a gradual change, but rather a rapid transition in kinematical properties around a gas fraction $\sim 20$ per cent.

Although our controlled experiments of Milky Way-sized galaxies highlight the impact of a specific parameter on the emerging stellar dynamics, other aspects like the disc growth, gas accretion, and galaxy mergers would also play important roles. Our conclusions thus call for further studies where the impact of such processes on a galaxy's AVR and orbital eccentricities can be quantified.

\section*{Acknowledgements}
We acknowledge support from the Knut and Alice Wallenberg Foundation and the Royal Physiographic Society of Lund. OA acknowledges support from the Swedish Research Council grant 2019-04659. Simulations were performed in part using computational resources at LUNARC, the centre for scientific and technical computing at Lund University, on the Swedish National Infrastructure for Computing (SNIC) allocation 2018/3-649, as well as allocation LU 2018/2-28 thanks to financial support from the Royal Physiographic Society of Lund. 

\section*{Data availability}
The data underlying this article will be shared on reasonable request to the corresponding author.

\bibliographystyle{mnras}
\bibliography{ms.bbl}

\begin{thebibliography}{}
\makeatletter
\relax
\def\mn@urlcharsother{\let\do\@makeother \do\$\do\&\do\#\do\^\do\_\do\%\do\~}
\def\mn@doi{\begingroup\mn@urlcharsother \@ifnextchar [ {\mn@doi@}
  {\mn@doi@[]}}
\def\mn@doi@[#1]#2{\def\@tempa{#1}\ifx\@tempa\@empty \href
  {http://dx.doi.org/#2} {doi:#2}\else \href {http://dx.doi.org/#2} {#1}\fi
  \endgroup}
\def\mn@eprint#1#2{\mn@eprint@#1:#2::\@nil}
\def\mn@eprint@arXiv#1{\href {http://arxiv.org/abs/#1} {{\tt arXiv:#1}}}
\def\mn@eprint@dblp#1{\href {http://dblp.uni-trier.de/rec/bibtex/#1.xml}
  {dblp:#1}}
\def\mn@eprint@#1:#2:#3:#4\@nil{\def\@tempa {#1}\def\@tempb {#2}\def\@tempc
  {#3}\ifx \@tempc \@empty \let \@tempc \@tempb \let \@tempb \@tempa \fi \ifx
  \@tempb \@empty \def\@tempb {arXiv}\fi \@ifundefined
  {mn@eprint@\@tempb}{\@tempb:\@tempc}{\expandafter \expandafter \csname
  mn@eprint@\@tempb\endcsname \expandafter{\@tempc}}}

\bibitem[\protect\citeauthoryear{{Abadi}, {Navarro}, {Steinmetz}  \&
  {Eke}}{{Abadi} et~al.}{2003}]{Abadi03a}
{Abadi} M.~G.,  {Navarro} J.~F.,  {Steinmetz} M.,   {Eke} V.~R.,  2003, \mn@doi
  [\apj] {10.1086/378316}, \href
  {http://adsabs.harvard.edu/abs/2003ApJ...597...21A} {597, 21}

\bibitem[\protect\citeauthoryear{{Agertz} \& {Kravtsov}}{{Agertz} \&
  {Kravtsov}}{2016}]{agertzkravtsov2016}
{Agertz} O.,  {Kravtsov} A.~V.,  2016, \mn@doi [\apj]
  {10.3847/0004-637X/824/2/79}, \href
  {http://adsabs.harvard.edu/abs/2016ApJ...824...79A} {824, 79}

\bibitem[\protect\citeauthoryear{{Agertz}, {Kravtsov}, {Leitner}  \&
  {Gnedin}}{{Agertz} et~al.}{2013}]{Agertz2013}
{Agertz} O.,  {Kravtsov} A.~V.,  {Leitner} S.~N.,   {Gnedin} N.~Y.,  2013,
  \mn@doi [\apj] {10.1088/0004-637X/770/1/25}, \href
  {http://adsabs.harvard.edu/abs/2013ApJ...770...25A} {770, 25}

\bibitem[\protect\citeauthoryear{{Agertz}, {Romeo}  \& {Grisdale}}{{Agertz}
  et~al.}{2015}]{AgertzRomeoGrisdale2015}
{Agertz} O.,  {Romeo} A.~B.,   {Grisdale} K.,  2015, \mn@doi [\mnras]
  {10.1093/mnras/stv440}, \href
  {http://adsabs.harvard.edu/abs/2015MNRAS.449.2156A} {449, 2156}

\bibitem[\protect\citeauthoryear{{Agertz} et~al.,}{{Agertz}
  et~al.}{2020}]{Agertz2020}
{Agertz} O.,  et~al., 2020, \mn@doi [\mnras] {10.1093/mnras/stz3053}, \href
  {https://ui.adsabs.harvard.edu/abs/2020MNRAS.491.1656A} {491, 1656}

\bibitem[\protect\citeauthoryear{{Agertz} et~al.,}{{Agertz}
  et~al.}{2021}]{Agertz2021}
{Agertz} O.,  et~al., 2021, \mn@doi [\mnras] {10.1093/mnras/stab322}, \href
  {https://ui.adsabs.harvard.edu/abs/2021MNRAS.503.5826A} {503, 5826}

\bibitem[\protect\citeauthoryear{{Aumer} \& {Binney}}{{Aumer} \&
  {Binney}}{2017}]{Aumer:2017aa}
{Aumer} M.,  {Binney} J.,  2017, \mn@doi [\mnras] {10.1093/mnras/stx1300},
  \href {https://ui.adsabs.harvard.edu/abs/2017MNRAS.470.2113A} {470, 2113}

\bibitem[\protect\citeauthoryear{{Aumer}, {Binney}  \& {Sch{\"o}nrich}}{{Aumer}
  et~al.}{2016a}]{Aumer:2016aa}
{Aumer} M.,  {Binney} J.,   {Sch{\"o}nrich} R.,  2016a, \mn@doi [\mnras]
  {10.1093/mnras/stw777}, \href
  {https://ui.adsabs.harvard.edu/abs/2016MNRAS.459.3326A} {459, 3326}

\bibitem[\protect\citeauthoryear{{Aumer}, {Binney}  \& {Sch{\"o}nrich}}{{Aumer}
  et~al.}{2016b}]{Aumer:2016ab}
{Aumer} M.,  {Binney} J.,   {Sch{\"o}nrich} R.,  2016b, \mn@doi [\mnras]
  {10.1093/mnras/stw1639}, \href
  {https://ui.adsabs.harvard.edu/abs/2016MNRAS.462.1697A} {462, 1697}

\bibitem[\protect\citeauthoryear{{Bensby}, {Alves-Brito}, {Oey}, {Yong}  \&
  {Mel{\'e}ndez}}{{Bensby} et~al.}{2011}]{Bensby2011}
{Bensby} T.,  {Alves-Brito} A.,  {Oey} M.~S.,  {Yong} D.,   {Mel{\'e}ndez} J.,
  2011, \mn@doi [\apjl] {10.1088/2041-8205/735/2/L46}, \href
  {https://ui.adsabs.harvard.edu/abs/2011ApJ...735L..46B} {735, L46}

\bibitem[\protect\citeauthoryear{{Bensby}, {Feltzing}  \& {Oey}}{{Bensby}
  et~al.}{2014}]{Bensby2014}
{Bensby} T.,  {Feltzing} S.,   {Oey} M.~S.,  2014, \mn@doi [\aap]
  {10.1051/0004-6361/201322631}, \href
  {https://ui.adsabs.harvard.edu/\#abs/2014A&A...562A..71B} {562, A71}

\bibitem[\protect\citeauthoryear{{Beraldo e Silva}, {Debattista}, {Nidever},
  {Amarante}  \& {Garver}}{{Beraldo e Silva} et~al.}{2021}]{Silva:2021aa}
{Beraldo e Silva} L.,  {Debattista} V.~P.,  {Nidever} D.,  {Amarante} J. A.~S.,
    {Garver} B.,  2021, \mn@doi [\mnras] {10.1093/mnras/staa3966}, \href
  {https://ui.adsabs.harvard.edu/abs/2021MNRAS.502..260B} {502, 260}

\bibitem[\protect\citeauthoryear{{Bird}, {Loebman}, {Weinberg}, {Brooks},
  {Quinn}  \& {Christensen}}{{Bird} et~al.}{2021}]{Bird:2021aa}
{Bird} J.~C.,  {Loebman} S.~R.,  {Weinberg} D.~H.,  {Brooks} A.~M.,  {Quinn}
  T.~R.,   {Christensen} C.~R.,  2021, \mn@doi [\mnras]
  {10.1093/mnras/stab289}, \href
  {https://ui.adsabs.harvard.edu/abs/2021MNRAS.503.1815B} {503, 1815}

\bibitem[\protect\citeauthoryear{{Bland-Hawthorn} \&
  {Gerhard}}{{Bland-Hawthorn} \& {Gerhard}}{2016}]{Bland:2016aa}
{Bland-Hawthorn} J.,  {Gerhard} O.,  2016, \mn@doi [\araa]
  {10.1146/annurev-astro-081915-023441}, \href
  {https://ui.adsabs.harvard.edu/abs/2016ARA&A..54..529B} {54, 529}

\bibitem[\protect\citeauthoryear{{Bournaud} \& {Elmegreen}}{{Bournaud} \&
  {Elmegreen}}{2009}]{Bournaud09}
{Bournaud} F.,  {Elmegreen} B.~G.,  2009, \mn@doi [\apjl]
  {10.1088/0004-637X/694/2/L158}, \href
  {http://adsabs.harvard.edu/abs/2009ApJ...694L.158B} {694, L158}

\bibitem[\protect\citeauthoryear{{Brook}, {Kawata}, {Gibson}  \&
  {Freeman}}{{Brook} et~al.}{2004}]{Brook:2004aa}
{Brook} C.~B.,  {Kawata} D.,  {Gibson} B.~K.,   {Freeman} K.~C.,  2004, \mn@doi
  [\apj] {10.1086/422709}, \href
  {https://ui.adsabs.harvard.edu/abs/2004ApJ...612..894B} {612, 894}

\bibitem[\protect\citeauthoryear{{Bryant} et~al.,}{{Bryant}
  et~al.}{2015}]{Bryant2015}
{Bryant} J.~J.,  et~al., 2015, \mn@doi [\mnras] {10.1093/mnras/stu2635}, \href
  {https://ui.adsabs.harvard.edu/abs/2015MNRAS.447.2857B} {447, 2857}

\bibitem[\protect\citeauthoryear{{Bundy} et~al.,}{{Bundy}
  et~al.}{2015}]{Bundy2015}
{Bundy} K.,  et~al., 2015, \mn@doi [\apj] {10.1088/0004-637X/798/1/7}, \href
  {https://ui.adsabs.harvard.edu/abs/2015ApJ...798....7B} {798, 7}

\bibitem[\protect\citeauthoryear{{Chabrier}}{{Chabrier}}{2003}]{chabrier03}
{Chabrier} G.,  2003, \mn@doi [\pasp] {10.1086/376392}, \href
  {http://adsabs.harvard.edu/abs/2003PASP..115..763C} {115, 763}

\bibitem[\protect\citeauthoryear{{Cioffi}, {McKee}  \& {Bertschinger}}{{Cioffi}
  et~al.}{1988}]{Cioffi1988}
{Cioffi} D.~F.,  {McKee} C.~F.,   {Bertschinger} E.,  1988, \mn@doi [\apj]
  {10.1086/166834}, \href {http://adsabs.harvard.edu/abs/1988ApJ...334..252C}
  {334, 252}

\bibitem[\protect\citeauthoryear{{Clarke} et~al.,}{{Clarke}
  et~al.}{2019}]{Clarke:2019aa}
{Clarke} A.~J.,  et~al., 2019, \mn@doi [\mnras] {10.1093/mnras/stz104}, \href
  {https://ui.adsabs.harvard.edu/abs/2019MNRAS.484.3476C} {484, 3476}

\bibitem[\protect\citeauthoryear{{Dessauges-Zavadsky}
  et~al.,}{{Dessauges-Zavadsky} et~al.}{2019}]{Dessauges:2019aa}
{Dessauges-Zavadsky} M.,  et~al., 2019, \mn@doi [Nature Astronomy]
  {10.1038/s41550-019-0874-0}, \href
  {https://ui.adsabs.harvard.edu/abs/2019NatAs...3.1115D} {3, 1115}

\bibitem[\protect\citeauthoryear{{Elmegreen}, {Elmegreen}, {Marcus},
  {Shahinyan}, {Yau}  \& {Petersen}}{{Elmegreen}
  et~al.}{2009}]{Elmegreen:2009aa}
{Elmegreen} D.~M.,  {Elmegreen} B.~G.,  {Marcus} M.~T.,  {Shahinyan} K.,  {Yau}
  A.,   {Petersen} M.,  2009, \mn@doi [\apj] {10.1088/0004-637X/701/1/306},
  \href {https://ui.adsabs.harvard.edu/abs/2009ApJ...701..306E} {701, 306}

\bibitem[\protect\citeauthoryear{{Feuillet}, {Frankel}, {Lind}, {Frinchaboy},
  {Garc{\'\i}a-Hern{\'a}ndez}, {Lane}, {Nitschelm}  \&
  {Roman-Lopes}}{{Feuillet} et~al.}{2019}]{Feuillet2019}
{Feuillet} D.~K.,  {Frankel} N.,  {Lind} K.,  {Frinchaboy} P.~M.,
  {Garc{\'\i}a-Hern{\'a}ndez} D.~A.,  {Lane} R.~R.,  {Nitschelm} C.,
  {Roman-Lopes} A.~r.,  2019, \mn@doi [\mnras] {10.1093/mnras/stz2221}, \href
  {https://ui.adsabs.harvard.edu/abs/2019MNRAS.489.1742F} {489, 1742}

\bibitem[\protect\citeauthoryear{{Frankel}, {Rix}, {Ting}, {Ness}  \&
  {Hogg}}{{Frankel} et~al.}{2018}]{Frankel2018}
{Frankel} N.,  {Rix} H.-W.,  {Ting} Y.-S.,  {Ness} M.,   {Hogg} D.~W.,  2018,
  \mn@doi [\apj] {10.3847/1538-4357/aadba5}, \href
  {https://ui.adsabs.harvard.edu/abs/2018ApJ...865...96F} {865, 96}

\bibitem[\protect\citeauthoryear{{Gaia Collaboration} et~al.,}{{Gaia
  Collaboration} et~al.}{2016}]{Gaia2016}
{Gaia Collaboration} et~al., 2016, \mn@doi [\aap]
  {10.1051/0004-6361/201629272}, \href
  {https://ui.adsabs.harvard.edu/abs/2016A&A...595A...1G} {595, A1}

\bibitem[\protect\citeauthoryear{{Genzel} et~al.,}{{Genzel}
  et~al.}{2015}]{Genzel2015}
{Genzel} R.,  et~al., 2015, \mn@doi [\apj] {10.1088/0004-637X/800/1/20}, \href
  {https://ui.adsabs.harvard.edu/abs/2015ApJ...800...20G} {800, 20}

\bibitem[\protect\citeauthoryear{{Glazebrook}}{{Glazebrook}}{2013}]{Glazebrook2013}
{Glazebrook} K.,  2013, \mn@doi [\pasa] {10.1017/pasa.2013.34}, \href
  {https://ui.adsabs.harvard.edu/abs/2013PASA...30...56G} {30, e056}

\bibitem[\protect\citeauthoryear{{Gnerucci} et~al.,}{{Gnerucci}
  et~al.}{2011}]{Gnerucci:2011aa}
{Gnerucci} A.,  et~al., 2011, \mn@doi [\aap] {10.1051/0004-6361/201015465},
  \href {https://ui.adsabs.harvard.edu/abs/2011A&A...528A..88G} {528, A88}

\bibitem[\protect\citeauthoryear{{Grand}, {Springel}, {G{\'o}mez}, {Marinacci},
  {Pakmor}, {Campbell}  \& {Jenkins}}{{Grand} et~al.}{2016}]{Grand:2016aa}
{Grand} R. J.~J.,  {Springel} V.,  {G{\'o}mez} F.~A.,  {Marinacci} F.,
  {Pakmor} R.,  {Campbell} D. J.~R.,   {Jenkins} A.,  2016, \mn@doi [\mnras]
  {10.1093/mnras/stw601}, \href
  {https://ui.adsabs.harvard.edu/abs/2016MNRAS.459..199G} {459, 199}

\bibitem[\protect\citeauthoryear{{Grisdale}, {Agertz}, {Romeo}, {Renaud}  \&
  {Read}}{{Grisdale} et~al.}{2017}]{Grisdale2017}
{Grisdale} K.,  {Agertz} O.,  {Romeo} A.~B.,  {Renaud} F.,   {Read} J.~I.,
  2017, \mn@doi [\mnras] {10.1093/mnras/stw3133}, \href
  {http://adsabs.harvard.edu/abs/2017MNRAS.466.1093G} {466, 1093}

\bibitem[\protect\citeauthoryear{{Grisdale}, {Agertz}, {Renaud}  \&
  {Romeo}}{{Grisdale} et~al.}{2018}]{Grisdale2018}
{Grisdale} K.,  {Agertz} O.,  {Renaud} F.,   {Romeo} A.~B.,  2018, \mn@doi
  [\mnras] {10.1093/mnras/sty1595}, \href
  {http://adsabs.harvard.edu/abs/2018MNRAS.tmp.1523G} {}

\bibitem[\protect\citeauthoryear{{Grisdale}, {Agertz}, {Renaud}, {Romeo},
  {Devriendt}  \& {Slyz}}{{Grisdale} et~al.}{2019}]{Grisdale2019}
{Grisdale} K.,  {Agertz} O.,  {Renaud} F.,  {Romeo} A.~B.,  {Devriendt} J.,
  {Slyz} A.,  2019, arXiv e-prints, \href
  {http://adsabs.harvard.edu/abs/2019arXiv190200518G} {}

\bibitem[\protect\citeauthoryear{{Guillet} \& {Teyssier}}{{Guillet} \&
  {Teyssier}}{2011}]{GuilletTeyssier2011}
{Guillet} T.,  {Teyssier} R.,  2011, \mn@doi [Journal of Computational Physics]
  {10.1016/j.jcp.2011.02.044}, \href
  {http://adsabs.harvard.edu/abs/2011JCoPh.230.4756G} {230, 4756}

\bibitem[\protect\citeauthoryear{{Gustafsson}, {Church}, {Davies}  \&
  {Rickman}}{{Gustafsson} et~al.}{2016}]{Gustafsson2016}
{Gustafsson} B.,  {Church} R.~P.,  {Davies} M.~B.,   {Rickman} H.,  2016,
  \mn@doi [\aap] {10.1051/0004-6361/201423916}, \href
  {https://ui.adsabs.harvard.edu/abs/2016A&A...593A..85G} {593, A85}

\bibitem[\protect\citeauthoryear{{Hayden} et~al.,}{{Hayden}
  et~al.}{2015}]{Hayden2015}
{Hayden} M.~R.,  et~al., 2015, \mn@doi [\apj] {10.1088/0004-637X/808/2/132},
  \href {http://adsabs.harvard.edu/abs/2015ApJ...808..132H} {808, 132}

\bibitem[\protect\citeauthoryear{{Hayden} et~al.,}{{Hayden}
  et~al.}{2020}]{Hayden2020}
{Hayden} M.~R.,  et~al., 2020, \mn@doi [\mnras] {10.1093/mnras/staa335}, \href
  {https://ui.adsabs.harvard.edu/abs/2020MNRAS.493.2952H} {493, 2952}

\bibitem[\protect\citeauthoryear{{Hayward} \& {Hopkins}}{{Hayward} \&
  {Hopkins}}{2017}]{HaywardHopkins2017}
{Hayward} C.~C.,  {Hopkins} P.~F.,  2017, \mn@doi [\mnras]
  {10.1093/mnras/stw2888}, \href
  {https://ui.adsabs.harvard.edu/abs/2017MNRAS.465.1682H} {465, 1682}

\bibitem[\protect\citeauthoryear{{Hernquist}}{{Hernquist}}{1990}]{Hernquist:1990aa}
{Hernquist} L.,  1990, \mn@doi [\apj] {10.1086/168845}, \href
  {https://ui.adsabs.harvard.edu/abs/1990ApJ...356..359H} {356, 359}

\bibitem[\protect\citeauthoryear{{Hernquist}}{{Hernquist}}{1993}]{Hernquist:1993aa}
{Hernquist} L.,  1993, \mn@doi [\apjs] {10.1086/191784}, \href
  {https://ui.adsabs.harvard.edu/abs/1993ApJS...86..389H} {86, 389}

\bibitem[\protect\citeauthoryear{{Hockney} \& {Eastwood}}{{Hockney} \&
  {Eastwood}}{1981}]{Hockney1981}
{Hockney} R.~W.,  {Eastwood} J.~W.,  1981, {Computer Simulation Using
  Particles}.
New York: McGraw-Hill, 1981

\bibitem[\protect\citeauthoryear{{Kassin} et~al.,}{{Kassin}
  et~al.}{2012}]{Kassin2012}
{Kassin} S.~A.,  et~al., 2012, \mn@doi [\apj] {10.1088/0004-637X/758/2/106},
  \href {http://adsabs.harvard.edu/abs/2012ApJ...758..106K} {758, 106}

\bibitem[\protect\citeauthoryear{{Kazantzidis}, {Zentner}, {Kravtsov},
  {Bullock}  \& {Debattista}}{{Kazantzidis} et~al.}{2009}]{Kazantzidis2009}
{Kazantzidis} S.,  {Zentner} A.~R.,  {Kravtsov} A.~V.,  {Bullock} J.~S.,
  {Debattista} V.~P.,  2009, \mn@doi [\apj] {10.1088/0004-637X/700/2/1896},
  \href {https://ui.adsabs.harvard.edu/abs/2009ApJ...700.1896K} {700, 1896}

\bibitem[\protect\citeauthoryear{{Khoperskov}, {Haywood}, {Snaith}, {Di
  Matteo}, {Lehnert}, {Vasiliev}, {Naroenkov}  \& {Berczik}}{{Khoperskov}
  et~al.}{2021}]{Khoperskov2021}
{Khoperskov} S.,  {Haywood} M.,  {Snaith} O.,  {Di Matteo} P.,  {Lehnert} M.,
  {Vasiliev} E.,  {Naroenkov} S.,   {Berczik} P.,  2021, \mn@doi [\mnras]
  {10.1093/mnras/staa3996}, \href
  {https://ui.adsabs.harvard.edu/abs/2021MNRAS.501.5176K} {501, 5176}

\bibitem[\protect\citeauthoryear{{Kim} \& {Ostriker}}{{Kim} \&
  {Ostriker}}{2014}]{KimOstriker2014}
{Kim} C.-G.,  {Ostriker} E.~C.,  2014, preprint, \href
  {http://adsabs.harvard.edu/abs/2014arXiv1410.1537K} {} (\mn@eprint {arXiv}
  {1410.1537})

\bibitem[\protect\citeauthoryear{{Kim} \& {Ostriker}}{{Kim} \&
  {Ostriker}}{2015}]{KimOstriker2015}
{Kim} C.-G.,  {Ostriker} E.~C.,  2015, \mn@doi [\apj]
  {10.1088/0004-637X/802/2/99}, \href
  {http://adsabs.harvard.edu/abs/2015ApJ...802...99K} {802, 99}

\bibitem[\protect\citeauthoryear{{Kim} et~al.,}{{Kim}
  et~al.}{2014}]{Kim:2014aa}
{Kim} J.-h.,  et~al., 2014, \mn@doi [\apjs] {10.1088/0067-0049/210/1/14}, \href
  {https://ui.adsabs.harvard.edu/abs/2014ApJS..210...14K} {210, 14}

\bibitem[\protect\citeauthoryear{{Kim} et~al.,}{{Kim}
  et~al.}{2016}]{Kim:2016aa}
{Kim} J.-h.,  et~al., 2016, \mn@doi [\apj] {10.3847/1538-4357/833/2/202}, \href
  {https://ui.adsabs.harvard.edu/abs/2016ApJ...833..202K} {833, 202}

\bibitem[\protect\citeauthoryear{{Krumholz} \& {Tan}}{{Krumholz} \&
  {Tan}}{2007}]{krumholztan07}
{Krumholz} M.~R.,  {Tan} J.~C.,  2007, \mn@doi [\apj] {10.1086/509101}, \href
  {http://adsabs.harvard.edu/abs/2007ApJ...654..304K} {654, 304}

\bibitem[\protect\citeauthoryear{{Lee}, {Miville-Desch{\^e}nes}  \&
  {Murray}}{{Lee} et~al.}{2016}]{Lee2016}
{Lee} E.~J.,  {Miville-Desch{\^e}nes} M.-A.,   {Murray} N.~W.,  2016, \mn@doi
  [\apj] {10.3847/1538-4357/833/2/229}, \href
  {https://ui.adsabs.harvard.edu/abs/2016ApJ...833..229L} {833, 229}

\bibitem[\protect\citeauthoryear{{Leitherer} et~al.,}{{Leitherer}
  et~al.}{1999}]{Leitherer1999}
{Leitherer} C.,  et~al., 1999, \mn@doi [\apjs] {10.1086/313233}, \href
  {http://adsabs.harvard.edu/abs/1999ApJS..123....3L} {123, 3}

\bibitem[\protect\citeauthoryear{{Li}, {Zhao}, {Zhai}  \& {Jia}}{{Li}
  et~al.}{2018}]{li:2018aa}
{Li} C.,  {Zhao} G.,  {Zhai} M.,   {Jia} Y.,  2018, \mn@doi [\apj]
  {10.3847/1538-4357/aac50f}, \href
  {https://ui.adsabs.harvard.edu/abs/2018ApJ...860...53L} {860, 53}

\bibitem[\protect\citeauthoryear{{Loebman}, {Ro{\v{s}}kar}, {Debattista},
  {Ivezi{\'c}}, {Quinn}  \& {Wadsley}}{{Loebman} et~al.}{2011}]{Loebman2011}
{Loebman} S.~R.,  {Ro{\v{s}}kar} R.,  {Debattista} V.~P.,  {Ivezi{\'c}}
  {\v{Z}}.,  {Quinn} T.~R.,   {Wadsley} J.,  2011, \mn@doi [\apj]
  {10.1088/0004-637X/737/1/8}, \href
  {https://ui.adsabs.harvard.edu/abs/2011ApJ...737....8L} {737, 8}

\bibitem[\protect\citeauthoryear{{Mackereth} et~al.,}{{Mackereth}
  et~al.}{2019}]{Mackereth:2019aa}
{Mackereth} J.~T.,  et~al., 2019, \mn@doi [\mnras] {10.1093/mnras/stz1521},
  \href {https://ui.adsabs.harvard.edu/abs/2019MNRAS.489..176M} {489, 176}

\bibitem[\protect\citeauthoryear{{Majewski} et~al.,}{{Majewski}
  et~al.}{2017}]{Apogee2017}
{Majewski} S.~R.,  et~al., 2017, \mn@doi [\aj] {10.3847/1538-3881/aa784d},
  \href {http://adsabs.harvard.edu/abs/2017AJ....154...94M} {154, 94}

\bibitem[\protect\citeauthoryear{{Martig}, {Minchev}  \& {Flynn}}{{Martig}
  et~al.}{2014}]{Martig:2014aa}
{Martig} M.,  {Minchev} I.,   {Flynn} C.,  2014, \mn@doi [\mnras]
  {10.1093/mnras/stu1322}, \href
  {https://ui.adsabs.harvard.edu/abs/2014MNRAS.443.2452M} {443, 2452}

\bibitem[\protect\citeauthoryear{{Martizzi}, {Faucher-Gigu{\`e}re}  \&
  {Quataert}}{{Martizzi} et~al.}{2015}]{Martizzi2015}
{Martizzi} D.,  {Faucher-Gigu{\`e}re} C.-A.,   {Quataert} E.,  2015, \mn@doi
  [\mnras] {10.1093/mnras/stv562}, \href
  {http://adsabs.harvard.edu/abs/2015MNRAS.450..504M} {450, 504}

\bibitem[\protect\citeauthoryear{{Mieda}, {Wright}, {Larkin}, {Armus},
  {Juneau}, {Salim}  \& {Murray}}{{Mieda} et~al.}{2016}]{Mieda:2016aa}
{Mieda} E.,  {Wright} S.~A.,  {Larkin} J.~E.,  {Armus} L.,  {Juneau} S.,
  {Salim} S.,   {Murray} N.,  2016, \mn@doi [\apj]
  {10.3847/0004-637X/831/1/78}, \href
  {https://ui.adsabs.harvard.edu/abs/2016ApJ...831...78M} {831, 78}

\bibitem[\protect\citeauthoryear{{Mikkola}, {McMillan}  \& {Hobbs}}{{Mikkola}
  et~al.}{2020}]{Mikkola2020}
{Mikkola} D.,  {McMillan} P.~J.,   {Hobbs} D.,  2020, \mn@doi [\mnras]
  {10.1093/mnras/staa1223}, \href
  {https://ui.adsabs.harvard.edu/abs/2020MNRAS.495.3295M} {495, 3295}

\bibitem[\protect\citeauthoryear{{Minchev}, {Famaey}, {Quillen}, {Dehnen},
  {Martig}  \& {Siebert}}{{Minchev} et~al.}{2012}]{Minchev2012}
{Minchev} I.,  {Famaey} B.,  {Quillen} A.~C.,  {Dehnen} W.,  {Martig} M.,
  {Siebert} A.,  2012, \mn@doi [\aap] {10.1051/0004-6361/201219714}, \href
  {https://ui.adsabs.harvard.edu/abs/2012A&A...548A.127M} {548, A127}

\bibitem[\protect\citeauthoryear{{Minchev}, {Chiappini}  \& {Martig}}{{Minchev}
  et~al.}{2013}]{Minchev2013}
{Minchev} I.,  {Chiappini} C.,   {Martig} M.,  2013, \mn@doi [\aap]
  {10.1051/0004-6361/201220189}, \href
  {https://ui.adsabs.harvard.edu/abs/2013A&A...558A...9M} {558, A9}

\bibitem[\protect\citeauthoryear{{Minchev} et~al.,}{{Minchev}
  et~al.}{2018}]{Minchev2018}
{Minchev} I.,  et~al., 2018, \mn@doi [\mnras] {10.1093/mnras/sty2033}, \href
  {https://ui.adsabs.harvard.edu/abs/2018MNRAS.481.1645M} {481, 1645}

\bibitem[\protect\citeauthoryear{{Murray}}{{Murray}}{2011}]{Murray2011b}
{Murray} N.,  2011, \mn@doi [\apj] {10.1088/0004-637X/729/2/133}, \href
  {http://adsabs.harvard.edu/abs/2011ApJ...729..133M} {729, 133}

\bibitem[\protect\citeauthoryear{{Navarro}, {Frenk}  \& {White}}{{Navarro}
  et~al.}{1996}]{Navarro:1996aa}
{Navarro} J.~F.,  {Frenk} C.~S.,   {White} S. D.~M.,  1996, \mn@doi [\apj]
  {10.1086/177173}, \href
  {https://ui.adsabs.harvard.edu/abs/1996ApJ...462..563N} {462, 563}

\bibitem[\protect\citeauthoryear{{Nordstr{\"o}m} et~al.,}{{Nordstr{\"o}m}
  et~al.}{2004}]{Nordstrom2004}
{Nordstr{\"o}m} B.,  et~al., 2004, \mn@doi [\aap] {10.1051/0004-6361:20035959},
  \href {https://ui.adsabs.harvard.edu/abs/2004A&A...418..989N} {418, 989}

\bibitem[\protect\citeauthoryear{{Quinn}, {Hernquist}  \& {Fullagar}}{{Quinn}
  et~al.}{1993}]{Quinn1993}
{Quinn} P.~J.,  {Hernquist} L.,   {Fullagar} D.~P.,  1993, \mn@doi [\apj]
  {10.1086/172184}, \href
  {https://ui.adsabs.harvard.edu/abs/1993ApJ...403...74Q} {403, 74}

\bibitem[\protect\citeauthoryear{Raiteri, Villata  \& Navarro}{Raiteri
  et~al.}{1996}]{Raiteri1996}
Raiteri C.~M.,  Villata M.,   Navarro J.~F.,  1996, A{\&}A, 315, 105

\bibitem[\protect\citeauthoryear{{Read}, {Lake}, {Agertz}  \&
  {Debattista}}{{Read} et~al.}{2008}]{Read2008}
{Read} J.~I.,  {Lake} G.,  {Agertz} O.,   {Debattista} V.~P.,  2008, \mn@doi
  [\mnras] {10.1111/j.1365-2966.2008.13643.x}, \href
  {http://adsabs.harvard.edu/abs/2008MNRAS.389.1041R} {389, 1041}

\bibitem[\protect\citeauthoryear{{Reddy}, {Tomkin}, {Lambert}  \& {Allende
  Prieto}}{{Reddy} et~al.}{2003}]{Reddy2003}
{Reddy} B.~E.,  {Tomkin} J.,  {Lambert} D.~L.,   {Allende Prieto} C.,  2003,
  \mn@doi [\mnras] {10.1046/j.1365-8711.2003.06305.x}, \href
  {https://ui.adsabs.harvard.edu/abs/2003MNRAS.340..304R} {340, 304}

\bibitem[\protect\citeauthoryear{{Renaud}, {Romeo}  \& {Agertz}}{{Renaud}
  et~al.}{2021a}]{Renaud2021}
{Renaud} F.,  {Romeo} A.~B.,   {Agertz} O.,  2021a, arXiv e-prints, \href
  {https://ui.adsabs.harvard.edu/abs/2021arXiv210600020R} {p. arXiv:2106.00020}

\bibitem[\protect\citeauthoryear{{Renaud}, {Agertz}, {Read}, {Ryde},
  {Andersson}, {Bensby}, {Rey}  \& {Feuillet}}{{Renaud}
  et~al.}{2021b}]{Renaud:2021aa}
{Renaud} F.,  {Agertz} O.,  {Read} J.~I.,  {Ryde} N.,  {Andersson} E.~P.,
  {Bensby} T.,  {Rey} M.~P.,   {Feuillet} D.~K.,  2021b, \mn@doi [\mnras]
  {10.1093/mnras/stab250}, \href
  {https://ui.adsabs.harvard.edu/abs/2021MNRAS.503.5846R} {503, 5846}

\bibitem[\protect\citeauthoryear{{Renaud}, {Agertz}, {Andersson}, {Read},
  {Ryde}, {Bensby}, {Rey}  \& {Feuillet}}{{Renaud} et~al.}{2021c}]{Renaud2021b}
{Renaud} F.,  {Agertz} O.,  {Andersson} E.~P.,  {Read} J.~I.,  {Ryde} N.,
  {Bensby} T.,  {Rey} M.~P.,   {Feuillet} D.~K.,  2021c, \mn@doi [\mnras]
  {10.1093/mnras/stab543}, \href
  {https://ui.adsabs.harvard.edu/abs/2021MNRAS.503.5868R} {503, 5868}

\bibitem[\protect\citeauthoryear{{Romeo}}{{Romeo}}{1992}]{romeo92}
{Romeo} A.~B.,  1992, \mnras, \href
  {http://adsabs.harvard.edu/abs/1992MNRAS.256..307R} {256, 307}

\bibitem[\protect\citeauthoryear{{Romeo} \& {Agertz}}{{Romeo} \&
  {Agertz}}{2014}]{Romeo2014}
{Romeo} A.~B.,  {Agertz} O.,  2014, \mn@doi [\mnras] {10.1093/mnras/stu954},
  \href {http://adsabs.harvard.edu/abs/2014MNRAS.442.1230R} {442, 1230}

\bibitem[\protect\citeauthoryear{{Romeo}, {Agertz}, {Moore}  \&
  {Stadel}}{{Romeo} et~al.}{2008}]{Romeo08}
{Romeo} A.~B.,  {Agertz} O.,  {Moore} B.,   {Stadel} J.,  2008, \mn@doi [\apj]
  {10.1086/591236}, \href {http://adsabs.harvard.edu/abs/2008ApJ...686....1R}
  {686, 1}

\bibitem[\protect\citeauthoryear{{Rosen} \& {Bregman}}{{Rosen} \&
  {Bregman}}{1995}]{rosenbregman95}
{Rosen} A.,  {Bregman} J.~N.,  1995, \mn@doi [\apj] {10.1086/175303}, \href
  {http://adsabs.harvard.edu/abs/1995ApJ...440..634R} {440, 634}

\bibitem[\protect\citeauthoryear{{Saintonge} et~al.,}{{Saintonge}
  et~al.}{2013}]{Saintonge:2013aa}
{Saintonge} A.,  et~al., 2013, \mn@doi [\apj] {10.1088/0004-637X/778/1/2},
  \href {https://ui.adsabs.harvard.edu/abs/2013ApJ...778....2S} {778, 2}

\bibitem[\protect\citeauthoryear{{Sales} et~al.,}{{Sales}
  et~al.}{2009}]{Sales:2009aa}
{Sales} L.~V.,  et~al., 2009, \mn@doi [\mnras]
  {10.1111/j.1745-3933.2009.00763.x}, \href
  {https://ui.adsabs.harvard.edu/abs/2009MNRAS.400L..61S} {400, L61}

\bibitem[\protect\citeauthoryear{{Santini} et~al.,}{{Santini}
  et~al.}{2014}]{Santini2014}
{Santini} P.,  et~al., 2014, \mn@doi [\aap] {10.1051/0004-6361/201322835},
  \href {https://ui.adsabs.harvard.edu/abs/2014A&A...562A..30S} {562, A30}

\bibitem[\protect\citeauthoryear{{Sch{\"o}nrich} \& {Binney}}{{Sch{\"o}nrich}
  \& {Binney}}{2009}]{SchonrichBinney2009}
{Sch{\"o}nrich} R.,  {Binney} J.,  2009, \mn@doi [\mnras]
  {10.1111/j.1365-2966.2009.15365.x}, \href
  {http://adsabs.harvard.edu/abs/2009MNRAS.399.1145S} {399, 1145}

\bibitem[\protect\citeauthoryear{{Schreiber} et~al.,}{{Schreiber}
  et~al.}{2015}]{Schreiber2015}
{Schreiber} C.,  et~al., 2015, \mn@doi [\aap] {10.1051/0004-6361/201425017},
  \href {https://ui.adsabs.harvard.edu/abs/2015A&A...575A..74S} {575, A74}

\bibitem[\protect\citeauthoryear{{Sellwood}}{{Sellwood}}{2014}]{Sellwood2014}
{Sellwood} J.~A.,  2014, \mn@doi [Reviews of Modern Physics]
  {10.1103/RevModPhys.86.1}, \href
  {https://ui.adsabs.harvard.edu/abs/2014RvMP...86....1S} {86, 1}

\bibitem[\protect\citeauthoryear{{Sellwood} \& {Binney}}{{Sellwood} \&
  {Binney}}{2002}]{SellwoodBinney2002}
{Sellwood} J.~A.,  {Binney} J.~J.,  2002, \mn@doi [\mnras]
  {10.1046/j.1365-8711.2002.05806.x}, \href
  {https://ui.adsabs.harvard.edu/abs/2002MNRAS.336..785S} {336, 785}

\bibitem[\protect\citeauthoryear{{Springel}}{{Springel}}{2000}]{Springel:2000aa}
{Springel} V.,  2000, \mn@doi [\mnras] {10.1046/j.1365-8711.2000.03187.x},
  \href {https://ui.adsabs.harvard.edu/abs/2000MNRAS.312..859S} {312, 859}

\bibitem[\protect\citeauthoryear{{Stern} et~al.,}{{Stern}
  et~al.}{2021}]{Stern2021}
{Stern} J.,  et~al., 2021, \mn@doi [\apj] {10.3847/1538-4357/abd776}, \href
  {https://ui.adsabs.harvard.edu/abs/2021ApJ...911...88S} {911, 88}

\bibitem[\protect\citeauthoryear{{Sutherland} \& {Dopita}}{{Sutherland} \&
  {Dopita}}{1993}]{sutherlanddopita93}
{Sutherland} R.~S.,  {Dopita} M.~A.,  1993, \mn@doi [\apjs] {10.1086/191823},
  \href {http://adsabs.harvard.edu/abs/1993ApJS...88..253S} {88, 253}

\bibitem[\protect\citeauthoryear{{Tacconi} et~al.,}{{Tacconi}
  et~al.}{2018}]{Tacconi2018}
{Tacconi} L.~J.,  et~al., 2018, \mn@doi [\apj] {10.3847/1538-4357/aaa4b4},
  \href {https://ui.adsabs.harvard.edu/abs/2018ApJ...853..179T} {853, 179}

\bibitem[\protect\citeauthoryear{{Teyssier}}{{Teyssier}}{2002}]{teyssier02}
{Teyssier} R.,  2002, \mn@doi [\aap] {10.1051/0004-6361:20011817}, \href
  {http://adsabs.harvard.edu/abs/2002A%26A...385..337T} {385, 337}

\bibitem[\protect\citeauthoryear{{Thornton}, {Gaudlitz}, {Janka}  \&
  {Steinmetz}}{{Thornton} et~al.}{1998}]{Thornton1998}
{Thornton} K.,  {Gaudlitz} M.,  {Janka} H.-T.,   {Steinmetz} M.,  1998, \mn@doi
  [\apj] {10.1086/305704}, \href
  {http://adsabs.harvard.edu/abs/1998ApJ...500...95T} {500, 95}

\bibitem[\protect\citeauthoryear{{Ting} \& {Rix}}{{Ting} \&
  {Rix}}{2019}]{Ting:2018aa}
{Ting} Y.-S.,  {Rix} H.-W.,  2019, \mn@doi [\apj] {10.3847/1538-4357/ab1ea5},
  \href {https://ui.adsabs.harvard.edu/abs/2019ApJ...878...21T} {878, 21}

\bibitem[\protect\citeauthoryear{{Wisnioski} et~al.,}{{Wisnioski}
  et~al.}{2015}]{Wisnioski:2015aa}
{Wisnioski} E.,  et~al., 2015, \mn@doi [\apj] {10.1088/0004-637X/799/2/209},
  \href {https://ui.adsabs.harvard.edu/abs/2015ApJ...799..209W} {799, 209}

\bibitem[\protect\citeauthoryear{{Yu} \& {Liu}}{{Yu} \&
  {Liu}}{2018}]{Yu:2018aa}
{Yu} J.,  {Liu} C.,  2018, \mn@doi [\mnras] {10.1093/mnras/stx3204}, \href
  {https://ui.adsabs.harvard.edu/abs/2018MNRAS.475.1093Y} {475, 1093}

\bibitem[\protect\citeauthoryear{{Yu} et~al.,}{{Yu} et~al.}{2021}]{Yu2021}
{Yu} S.,  et~al., 2021, \mn@doi [\mnras] {10.1093/mnras/stab1339}, \href
  {https://ui.adsabs.harvard.edu/abs/2021MNRAS.505..889Y} {505, 889}

\bibitem[\protect\citeauthoryear{{de Jong} et~al.,}{{de Jong}
  et~al.}{2019}]{4MOST2019}
{de Jong} R.~S.,  et~al., 2019, \mn@doi [The Messenger]
  {10.18727/0722-6691/5117}, \href
  {https://ui.adsabs.harvard.edu/abs/2019Msngr.175....3D} {175, 3}

\bibitem[\protect\citeauthoryear{{van Dokkum} et~al.,}{{van Dokkum}
  et~al.}{2013}]{Dokkum:2013aa}
{van Dokkum} P.~G.,  et~al., 2013, \mn@doi [\apjl]
  {10.1088/2041-8205/771/2/L35}, \href
  {https://ui.adsabs.harvard.edu/abs/2013ApJ...771L..35V} {771, L35}

\makeatother
\end{thebibliography}


\bsp	
\label{lastpage}
\end{document}